# Axial Magnetic Field Effect on Vacuum Arc Thruster Performance

**IEPC-2019-662**

Igal Kronhaus[1], Satyajit Chowdhury[2] and Matteo Laterza[3]
*Faculty of Aerospace Engineering, Technion - Israel Institute of Technology, Haifa, 3200003, Israel*

**Abstract:** The performance gain of co-axial vacuum arc thruster in external axial magnetic field is analyzed by power and direct thrust measurements. Thruster geometry is actively controlled and inspected using laser profilometry. Peak performance is shown to be obtained in magnetic induction of 0.1 - 0.2 T, with measured thrust of 21 μN at arc power of 2.7 W, and thruster efficiency of ≈ 4 %. The performance gain is attributed to ion beam focusing, increased erosion rate, and ion acceleration due to the axial magnetic field.

## Nomenclature

| | | |
|---|---|---|
| $B$ | = | Axial magnetic induction |
| $C_t$ | = | Thrust correction factor |
| $E_r$ | = | Cathode erosion rate |
| $f$ | = | Arc pulse firing frequency |
| $I_{sp}$ | = | Specific impulse |
| $\dot{m}$ | = | Cathode consumption rate |
| $P_{arc}$ | = | Arc power |
| $P_{in}$ | = | Input power |
| $T$ | = | Thrust |
| $\epsilon_p$ | = | Arc pulse energy |
| $\eta_e$ | = | Electrical |
| $\eta_{thruster}$ | = | Thruster efficiency |
| $\eta_{tot}$ | = | Total efficiency |

## I. Introduction

Vacuum arc thrusters (VAT) are promising propulsion devices for nano-satellites and CubeSats,[1,2] with a few examples experimentally operated in space. VAT are pulsed-dc devices that utilize arc discharge, across an insulator, between two electrodes to produce thrust. The main advantages of the VAT compared to other electric propulsion devices is in its simplicity and scalability to very low power (< 10 W). In a VAT the cathode electrode is consumed as propellant during the discharge. The cathode is eroded in localized regions, where the discharge is attached, known as cathodic spots.[3] The metal plasma emitted from these micrometer sized spots is naturally accelerated by gas-dynamic expansion.[4] Further acceleration/focusing of the plasma plume can be achieved by the use of an axial magnetic field [5]. Significant technical development of these devices was made by several groups in the last decade. There are two established configuration: co-axial and the ring shaped,[6] termed according to the shape and placement of the cathode with respect to the anode. As any rocket it is important maintain controlled feeding of the propellant. However, all these VAT designs either have fixed geometry or use passive feeding mechanisms. Recently a co-axial VAT was developed with an active, fully controlled, feeding mechanism known as the inline-screw-feeding VAT (ISF-VAT).[7,8] By utilizing the ISF-VAT technology it is possible to accurately and reliably evaluate the performance enhancement of co-axial with an external axial magnetic field.

---
[1] Head, Aerospace Plasma Laboratory, kronhaus@technion.ac.il
[2] Postdoctoral Fellow, Aerospace Plasma Laboratory
[3] Postdoctoral Fellow, Aerospace Plasma Laboratory

## II. Experimental Setup

### A. ISF-VAT with Magnetic Coil

The ISF-VAT concept[7] is shown in Fig. 1. It is a co-axial vacuum arc thruster with an active feeding mechanism. A central cathode rod is freely disposed within a concentric insulator tube. A second electrode, positioned at the outer edge of the insulator, functions both as the anode of the dc circuit and as the exit plane of the thruster. To keep the VAT geometry constant during long duration operations, the cathode, connected to a metallic headless screw, is advanced at a precise rate inside the insulator in a helical path. The screw provides also an electrical contact with the thruster body that is under negative potential. With the correct selection of the linear advance rate and screw pitch, a balance between cathode erosion and feeding can be achieved. The helical motion both compensates for the radial as well as the azimuthal cathode erosion patterns. This allows for mainlining near constant thruster geometry throughout the operational life of the thruster as well as improved uniformity of the recoating process, i.e. the process of replenishing the conducting layer on the cathode-insulator-anode interface. To keep power consumption and thruster dimensions to a minimum a mechanism comprised of a spiral spring and an amplified-piezoelectric-actuator (APB) was chosen to power the cathode rotation, as shown in Fig. 1(a). The mechanical energy is stored in the spring, and the APB regulates the spring unwinding motion.

Following earlier attempts at magnetic enhancement of VATs, see for example the review in Ref. 6, a simple air coil arrangement wound around the ISF-VAT anode was selected to evaluate the effect of magnetic axial field on thruster performance. The coil is made from 45 turns of 24 AWG copper with inner diameter of 18 mm and is 10 mm in length. The coil is held on the anode by its Teflon bobbin. To achieve maximum magnetic induction on the cathode surface the coil mid length is position at the cathode-insulator-anode surface plane. The calculated magnetic field and its relative position with respect to the cathode are shown in Fig. 1(b).

### B. Propulsion Module and Measurement Setup

The experiments were performed on an ISF-VAT propulsion module (PM)[8] being developed for the DriveSat CubeSat[9] at the aerospace plasma laboratory (APL), Technion. The PM includes two ISF-VATs, a power processing unit (PPU), and a structural frame to interface the module to the CubeSat bus. The PM volume is 2 cm × 9 cm × 9 cm with "wet" mass of 200 g. The PM's PPU, in addition to discharge power, provides control and power to the active feeding mechanism. Figure 2(a) shows a photograph of the laboratory PM. This PM contains only one ISF-VAT and has a modified frame for easy interference with APL's vacuum chamber.

The vacuum chamber is 1.2 m long and 0.6 m in diameter. Two $700 \; l/s$ turbo molecular pumps backed by $300 \; l/m$ rotary vane pumps are used to maintain a base pressure $\sim 10^{-6}$ mbar. The chamber pressure is measured by a Pirani gauge. The PM is placed on a micro-Newton resolution torsion thrust balance,[10] as shown in Fig. 2(b). The PM is powered by 16 V dc provided by a laboratory power supply outside the chamber. The PM is monitored via serial communication to a PC. In addition to discharge circuit onboard PPU[8], the coil is powered by a dedicated magnetic circuit, placed outside the chamber that is synchronized with the thruster firing pulses and allows regulation of the coil current. A schematic of the setup is shown in Fig. 2(c).

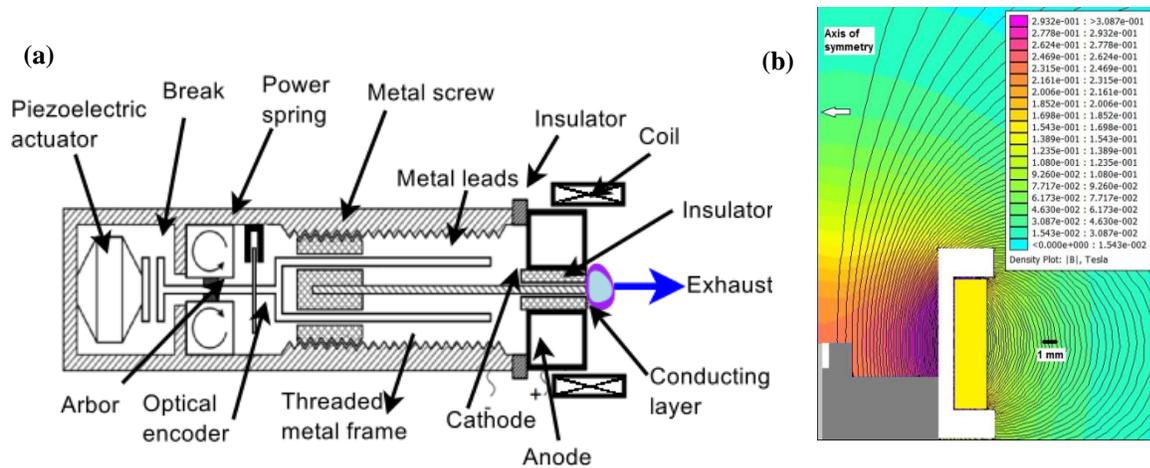

**Figure 1.** Schematics of the ISF-VAT with added magnetic coil (a) and magnetic simulation at 0.25 T (b), the cathode is positioned in the axis of symmetry of the simulation domain.

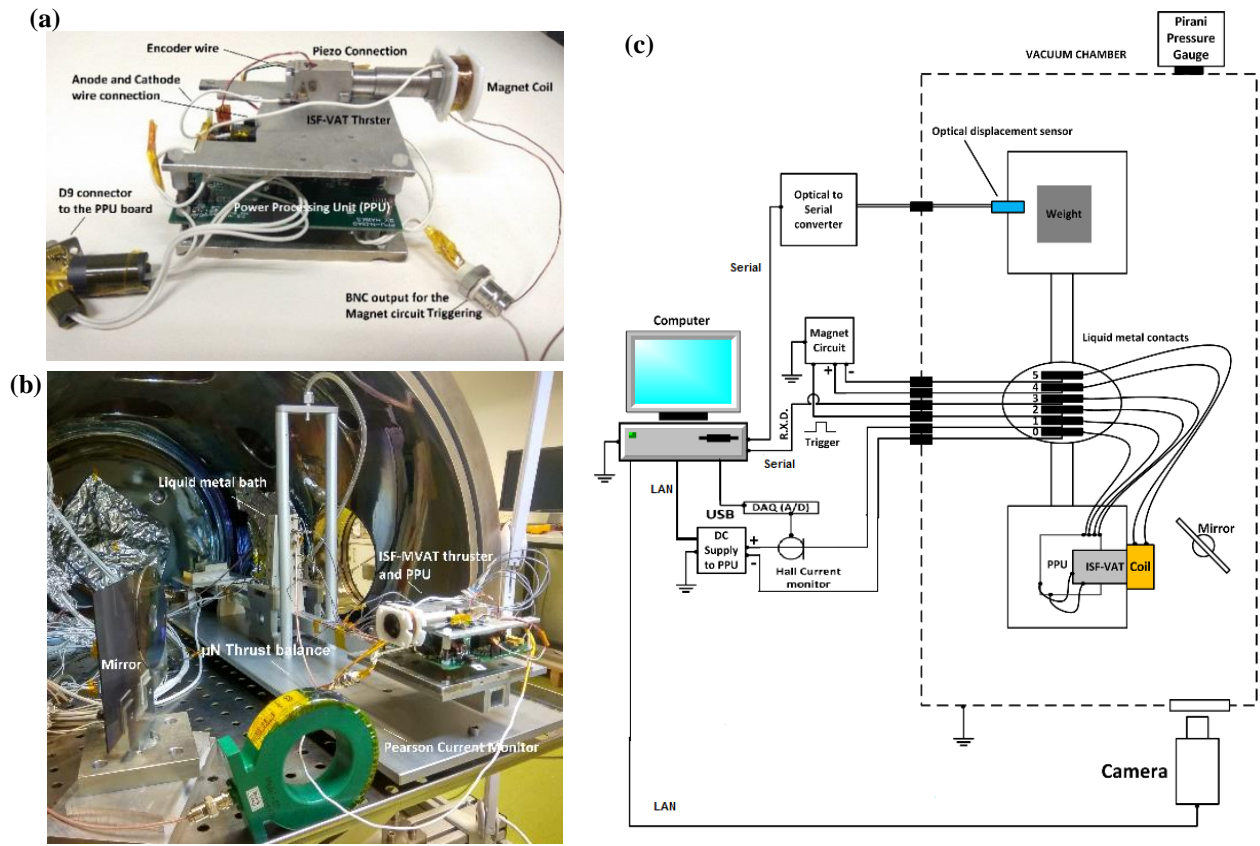

**Figure 2.** Photograth of the PM (a) and PM installation on the thrust balance (b) and a schematic of the experimental setup (c).

To enable automated testing and fault detection, such as a thruster shot-circuit, a power protection system based on monitoring the PM/PPU input current was built. This setup, shown in Fig. 2(c), is comprised of a Hall current monitor, a DAQ for analog to digital conversion, a remote controlled power supply, and a PC running LabVIEW. In addition to power protection, the setup is capable of continuous measurement of PM input current. The logging of input current data is synchronized with the thrust balance data (sampling rate of 2 Hz). In addition to the PM input current, high temporal resolution measurements of the arc current and arc voltage were manually performed using a Pearson current monitor (inside the chamber) and a differential high voltage probe (outside the chamber) recorded on a 5 Gs/s oscilloscope. The magnet coil current was also measured, using a Hall current monitor, by the oscilloscope. Video recording of the plume was performed in each experiment using a charged coupled device (CCD) color camera. Before operating the thruster inside the vacuum chamber, the thruster cathode-insulator-anode interface was coated by graphite. This operation is performed only once for a specific combination of cathode and insulator. The same thruster anode was used throughout the entire campaign.

## III. Experimental Results

### A. Single Arc Pulse Current-Voltage Characteristics

Time depended current-voltage measurements of single arc pulses were conducted at several magnetic induction values. These experiments were conducted without feeding. The cathode was manually set to position and then the thruster was operated at arc pulse frequency of 15 Hz for duration of 90 s. At the end of the firing cycle the thruster was removed from the chamber and the cathode repositioned. Two examples are shown in Fig. 3. We observe that the arc voltage is ~ 30 V and is independent of the magnetic field strength. The discharge, however, is suppressed at $B > 0.2$ T, operating for shorter arc pulse widths. Individual I-V curves were integrated to obtain a characteristic of the arc pulse energy versus magnetic induction; the results are shown in Fig. 4. We observe a rapid increase in arc pulse energy $B \approx 0 - 0.05$ T, maintaining $\varepsilon_p \approx 150$ mJ - 175 mJ $B \approx 0.07 - 0.2$ T, at greater magnetic induction the arc pulse energy is significantly reduced.

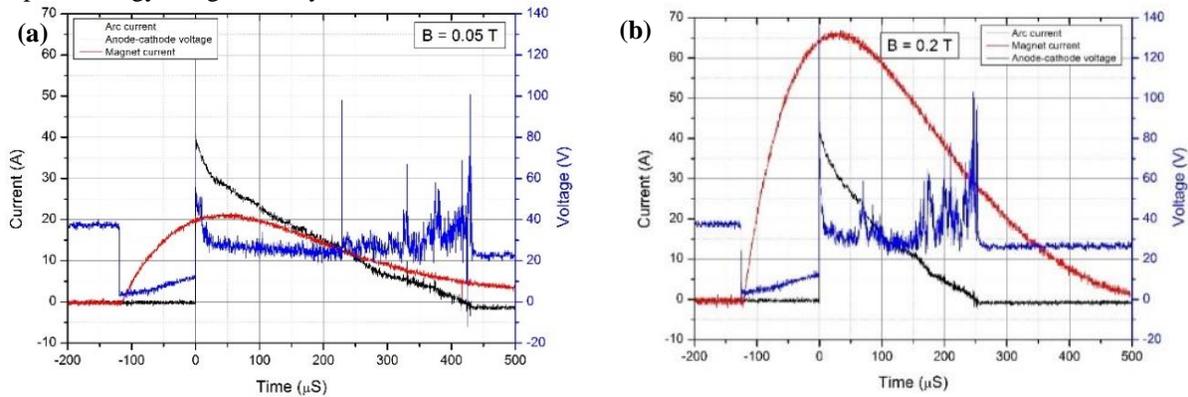

**Figure 3. Arc I-V and magnet current versus time for a single pulse at $B = 0.05$ T (a), $B = 0.2$ T (b).**

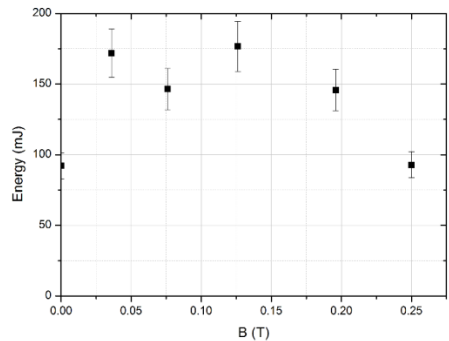

**Figure 4. Measured single arc pulse energy $\varepsilon_p$ versus magnetic induction.**

## B. Thruster Performance

Thrust and power measurements were conducted without feeding. The cathode was manually set to position and then the thruster was operated at arc pulse frequency of 15 Hz for duration of 90 s. At the end of the firing cycle the thruster was removed from the chamber and the cathode repositioned. Typical thrust measurements at several magnetic induction values are shown in Fig. 5. We observe that in most cases higher thrust is obtained at the initial operation and then gradually decreases until the firing is terminated, this is caused by the cathode material being consumed, and is the reason why feeding is necessary for long duration operation.

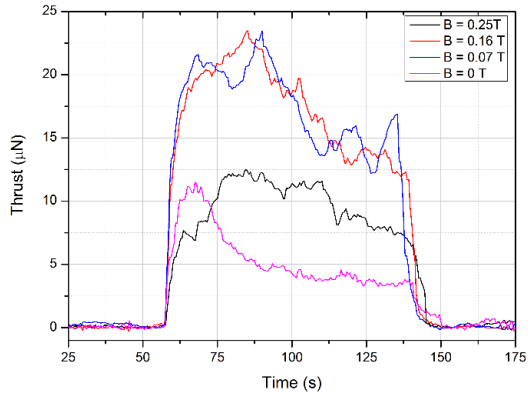

**Figure 5. Measured thrust versus time at several magnetic induction. Sample integration time is 0.5 s.**

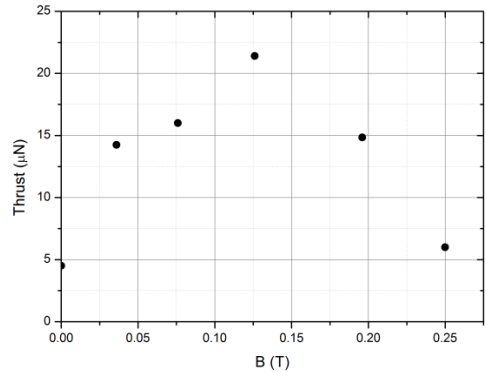

**Figure 6. Average thrust versus magnetic induction.**

These measurements were time averaged over the firing cycle to produce characteristics of thrust versus magnetic induction, shown in Fig. 6. We observe the existence of a distinct maximum in thrust at $B = 0.12$ T with substantially lower values measured at either lower or higher magnetic induction; we note however that the greatest increase in thrust is in the range $B \approx 0 - 0.04$ T, see the discussion in section IV. Each of the thrust measurements was accompanied by a precise measurement of the cathode (Ti) mass consumption. This was evaluated by comparing cathode topology before and after firing measured using laser profilometry. The technique is described in detail in Ref. 11. The cathode consumed mass divided by the firing time provides the average mass consumption rate $\dot{m}$, shown in Fig. 7. We observe a maximum in the mass consumption rate at $B = 0.12$ T, doubling $\dot{m}$ compared to the non-magnetic case.

The availability of data on both arc current and mass consumption allows us to evaluate the erosion rate coefficient $E_r$ variation with magnetic induction, as shown in Fig. 8. The erosion rate increases with the magnetic field until a maximum is reached at $B \approx 0.12$ T where $E_r \approx 31$ µg/C is very close to the literature value.[12] The erosion rate sharply drops for $B > 0.2$ T.

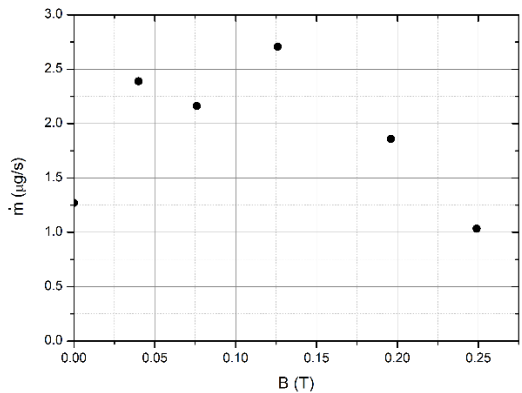

**Figure 7. Cathode mass consumption rate versus magnetic induction.**

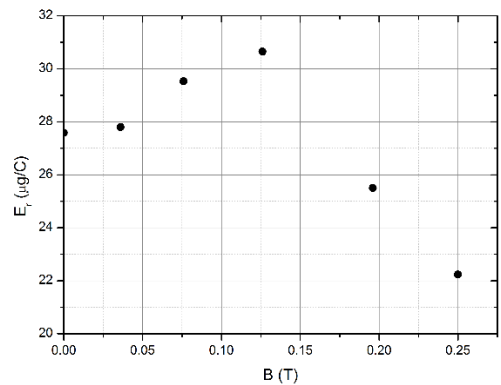

**Figure 8. Cathode erosion rate versus magnetic induction.**

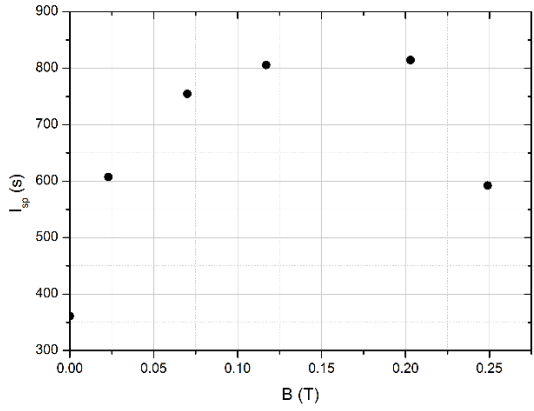 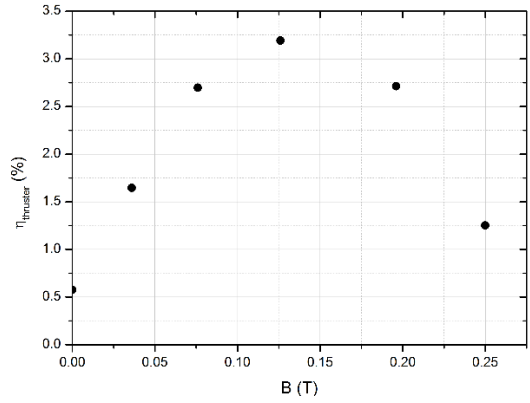

**Figure 9. Thruster specific impulse versus magnetic induction.**   **Figure 10. Thruster efficiency versus magnetic induction.**

Additional thruster performance parameters can be computed. Figure 9 shows the thruster specific impulse defined as:

$$I_{sp} = \frac{T}{\dot{m} g_0} \qquad (1)$$

where $T$ is the thrust and $g_0 = 9.81$ m/s². The thruster efficiency, shown in Fig. 10, is defined as:

$$\eta_{thruster} = \frac{T^2}{2\dot{m} P_{arc}} \qquad (2)$$

where the average arc power is computed by $P_{arc} = f \varepsilon_p$ and $f = 15$ Hz is the arc pulse firing frequency. From these results an optimum thruster performance is obtained at $B \approx 0.12$ T.

### C. Propulsion Module Performance

Long duration operation was characterized by operating the PM as designed with a repetitive firing/feeding cycle. The PM/PPU was programmed to feed the cathode at equivalent rate of $\dot{m} = 1.3$ µg/s between firing. The firing cycle duration was 90 sec with constant firing frequency of 15 Hz. A cool down time of 300 s is programmed between firings. During these tests PM power supply was set 16 V and both the input current and arc current were monitored. The input current and thrust data were recorded simultaneously at a sampling rate of 2 Hz. A typical data set of continues firing sequence is shown in Fig. 11.

The PPU electrical efficiency defined as:

$$\eta_e = \frac{P_{arc}}{P_{in}} \qquad (3)$$

where $P_{in}$ in the PM /PPU input power. As shown in Fig. 12, $\eta_e$ is obtained by a fitting the power data. At the power levels of interest the $\eta_e \approx 30 - 40$ %. This is a relatively modest efficiency for this type of PPU and can be potentially improved to ~ 90 %. We note that the magnetic circuit power is not included in the PM power budget.

The PM average thrust and input power during operation is presented in Fig. 13. The data is averaged over several firing sequences for each magnetic induction value. We observe here the thrust do not exceed 15 µN, this is due to active feeding mechanism that imposes $\dot{m} = 1.3$ µg/s, effectively "starving" the discharge. This in contrast to the results in section III.B were the mass consumption was established freely by the discharge. Figure 14 presents the PM's total efficiency, defined as:

$$\eta_{tot} = \frac{T^2}{2\dot{m} P_{in}} = \eta_e \eta_{thruster}. \qquad (4)$$

Figure 14 also presented the thruster efficiency as calculated according to $\eta_{thruster} = \eta_{tot}/\eta_e$, where $\eta_e$ is known from Fig. 12. We calculate $\eta_{thruster} = 3.5$ % at $B = 0.12$ T which is similar to the measured value in the data set of section III.B.

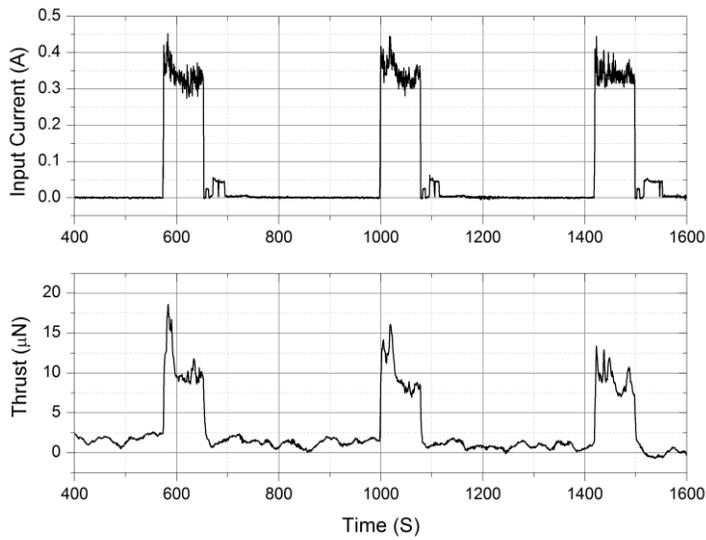
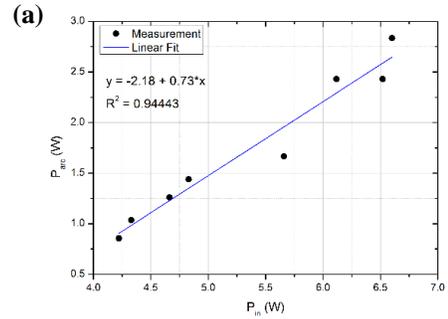
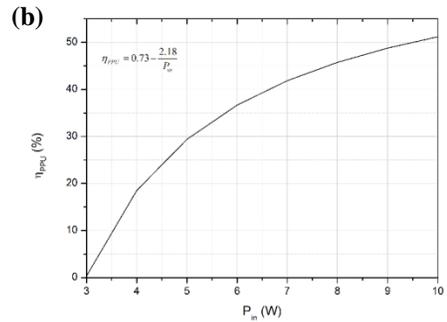

**Figure 11. Current and thrust data during PM/PPU operation. Data shows 3 separate firing sequences, including feeding and cool down time between firings. 0.5 s integration time.**

**Figure 12. PM/PPU electrical efficiency. Arc power (a) and electrical efficiency (b) versus input power.**

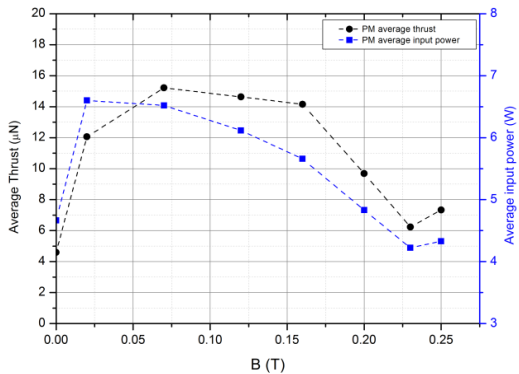
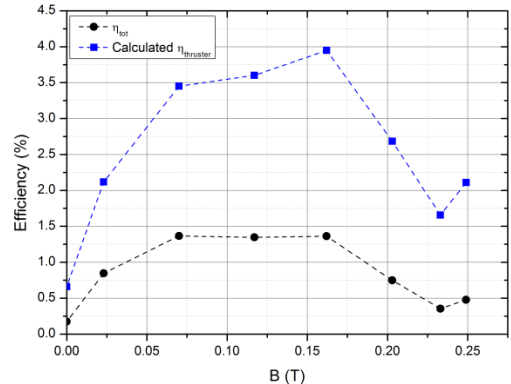

**Figure 13. Averaged thrust and PM input power versus magnetic induction.**

**Figure 14. PPU electrical efficiency and computed thruster efficiency versus magnetic induction.**

## IV. Discussion

### A. Plume Shape and Thrust Correction Factor

When evaluating a plasma thruster performance one has to consider beam divergence, i.e. the ejected ion spatial distribution. Assuming a cosine distribution of the plume and coplanar cathode, insulator and anode surfaces, then without coil there is no shadowing of the plasma flux and the thrust correction factor is $C_t^p = 0.67$,[12] where $C_t^p = 1$ is a perfect beam collimation. A photograph of such a plume, i.e. the ISF-VAT plume, is shown in Fig. 15(a). When assembling the coil there is an additional shadowing effect of the plume that reduces the thrust. Of plume for the thruster with magnetic coil were captured at several applied magnetic field induction values, and are shown in Fig.

15. As the applied magnetic field increases in magnitude the plume becomes more beam like and compressed towards the axis with and less visible interaction with the coil inner wall.

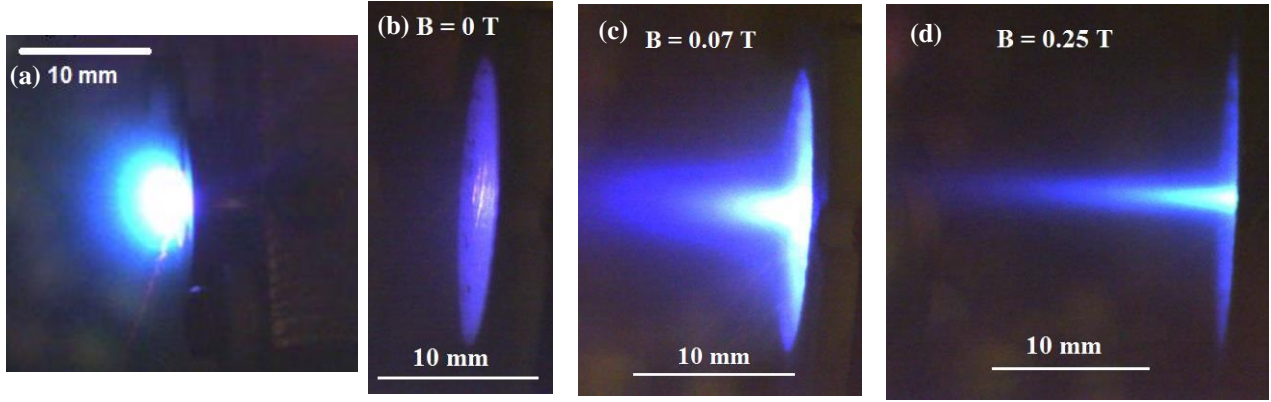

**Figure 15. Photograph of the plasma plume for no-coil (a) unpowered coil (b) 0.07 T (c) and 0.25 T (d). All images use 1 ms exposure time.**

Because the magnetic coil is protruding outside the anode, as shown in Fig. 16, there is an additional shadowing effect of the plume. The thrust correction factor due to the coil $C_t^{coil}$ can be evaluated by the line of sight angle $\varphi$ as:

$$C_t^{coil} = \frac{\int_0^{\pi/2-\varphi_0} \cos\varphi \, d\varphi}{\int_0^{\pi/2} \cos\varphi \, d\varphi} = 0.78 \qquad (3)$$

where $\varphi_0 = \tan^{-1}(L/r_a)$, $r_{coil} = 9$ mm, and $L = 7$ mm. The total thrust correction factor is therefore:

$$C_{tot} = C_t^p C_t^{coil} = 0.52 \qquad (4)$$

These results are summarized in Table 1. We can conclude that in addition to focusing, other mechanisms such as increased erosion rate and ion acceleration contribute to performance enhancement of the ISF-VAT in axial magnetic field.

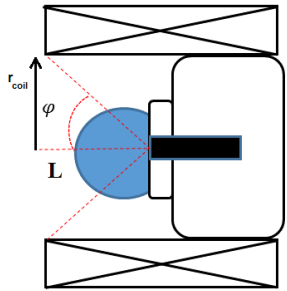

**Figure 16. Thruster geometry showing the relative position of the coil and anode.**

**Table 1. Comparison between measured and geometrically corrected (with respect to no coil) $T/P_{arc}$ results.**

| Case | Thrust Corrected Factor $C_{tot}$ | Measured $T/P_{arc}$, µN/W | Corrected $T/P_{arc}$, µN/W |
|---|---|---|---|
| No coil | 0.67 | 4.3 | - |
| Unpowered coil (0 T) | 0.52 | 3.3 | 3.3 |
| Powered coil (0.07 T) | 1.0 | 7.3 | 6.4 |
| Powered coil (0.12 T) | 1.0 | 8.0 | 6.4 |

## Conclusions

The performance enhancement over non-magnetic ISF-VAT due to application of external axial magnetic field is shown to be substantial, doubling the thrust to power ratio and enabling thrust generation of 21 μN at 2.7 W of average arc power, with 4 % thruster efficiency. The maximum performance is achieved at magnetic induction in the range 0.1 - 0.2 T. At higher magnetic induction the discharge is suppressed. The performance gain is attributed to not only ion beam focusing, but also increased erosion rate and ion acceleration. The current power processing unit efficiency is shown to be about ≈ 40 %. Future work includes design of a permanent magnet system and improvement of PPU efficiency.

## Acknowledgments

This work was funded by the Israeli Ministry of Science, Technology and Space.